\documentclass[preprint,showpacs,amsmath,amssymb,aps,prd,nofootinbib]{revtex4}
\usepackage{epsfig,graphicx,color,appendix}
\begin{document}
\begin{flushright}
IBS-CTPU-16-10
\end{flushright}
\title{\mbox{}\\[10pt]
QCD Axion as a Bridge \\Between String Theory and Flavor Physics}

\author{Y. H. Ahn}
\email{yhahn@ibs.re.kr}
\affiliation{Center for Theoretical Physics of the Universe, Institute for Basic Science (IBS), Daejeon, 34051, Korea}


\begin{abstract}
\noindent We construct a string-inspired model, motivated by the flavored Peccei-Quinn (PQ) axions, as a useful bridge between flavor physics and string theory. The key feature is two anomalous gauged $U(1)$ symmetries, responsible for both the fermion mass hierarchy problem of the standard model and the strong CP problem, that combine string theory with flavor physics and severely constrain the form of the F- and D-term contributions to the potential. In the context of supersymmetric moduli stabilization we stabilize the size moduli with positive masses while leaving two axions massless and one axion massive. We demonstrate that, while the massive gauge bosons eat the two axionic degrees of freedom, two axionic directions survive to low energies as the flavored PQ axions.
\end{abstract}
\maketitle %
%%%%%%%%%%%%%%%%%%%%%%%%%%%%%%%%
\section{Introduction}
The standard model (SM) as an effective theory has been successful in describing phenomena until now, but it suffers from theoretical problems (inclusion of gravity in gauge theory, instability of the Higgs potential, the SM fermion mass hierarchies and their mixing patterns with the CP violating phases, the strong CP problem~\cite{Peccei:1977hh}, etc) and cosmological issues (dark matter, inflation, cosmological constant, etc). It is widely believed that the SM should be extended to a more fundamental underlying theory. If nature is stringy, string theory should give insight into all such fundamental problems. Therefore, we can anticipate that there may exist some correlation between string theory as a fundamental theory and low energy flavor physics.

Ref.~\cite{Ahn:2014gva} used a superpotential for unifying flavor and strong CP problems, the so-called flavored Peccei-Quinn (PQ) symmetry model, in a way that no axionic domain wall problem occurs. 
In this Letter we construct an explicit string-inspired model, motivated by the flavored PQ axions, as a useful bridge between flavor physics and string theory. The key features of the model can be present in type IIB compactification. The crucial one is two anomalous gauged $U(1)$ symmetries that combine string theory with flavor physics, and severely constrain the form of the F- and D-term contributions to the potential. We show how supersymmetric moduli stabilization with three fixed size moduli, one fixed axionic partner and  two unfixed axions can be realized. And we illustrate that the model admits metastable vacuum with spontaneously broken supersymmetry (SUSY) and a nearly vanishing positive vacuum energy, resulting from the positive contributions to the potential associated with the gauge symmetry of the theory, the so-called D-terms. In addition, we illustrate how to achieve phenomenologically non-trivial vacuum expectation value (VEV) directions of flavon fields.  Finally, we demonstrate that, while the massive gauge bosons eat the axionic degree of freedoms, two axionic directions survive to low energies as the flavored PQ axions~\cite{Ahn:2014gva}.

%%%%%%%%%%%%%%%%%%%%%%%%%%%%%%%%
\section{The model}
Below the scale where the dilation and complex structure moduli are stabilized through fluxes~\cite{Gukov:1999ya}, we consider the low energy Kahler potential $K$ and superpotential $W$ for the Kahler moduli and matter superfields invariant under gauged $U(1)_X$ symmetry
%\begin{widetext}
 \begin{eqnarray}
  K&=&-M^2_P\ln\left\{(T+\bar{T})\prod^2_{i=1}\left(T_i+\bar{T}_i-\frac{\delta^{\rm GS}_i}{16\pi^2}V_{X_i}\right)\right\}
  +\sum^2_{i=1}Z_i\Phi^\dag_i e^{-X_iV_{X_i}}\Phi_i\nonumber\\
  &+&\sum_k Z_k|\varphi_k|^2+...\label{Kahler0}\\
  W&=& W_{Y}+W_v+W_0+W(T)
 \label{Kahler}
 \end{eqnarray}
%\end{widetext}
which is appropriate for toroidal orientifold, where $M_P=m_{\rm P}/\sqrt{8\pi}=2.4\times10^{18}$ GeV is the reduced Planck mass, and dots stand for higher order terms.  The first term in Eq.~(\ref{Kahler0}) has a no-scale symmetry up to perturbative corrections from string theory. Note that the Kahler moduli do not appear in the superpotential at tree level, therefore they are not fixed by the fluxes. From the Kahler potential and superpotential, we schematically obtain the low-energy effective Lagrangian 
 \begin{eqnarray}
  {\cal L}\supset
 \label{eff_Lag}\frac{1}{2}K_{T\bar{T}}\partial_{\mu}T\partial^{\mu}\bar{T}+\frac{1}{2}K_{T_i\bar{T}_i}\partial_{\mu}T_i\partial^{\mu}\bar{T}_i-V+{\cal L}(\Phi_i,\varphi_i,...)\,.
  \label{Kahler_metric}
 \end{eqnarray}
Here the kinetic terms for the axionic and size moduli do not mix in perturbation theory, due to the axionic shift symmetry, where any nonperturbative violations are small enough to be irrelevant.
And the Kahler metric in Eq.~(\ref{Kahler_metric}) is given by
 \begin{eqnarray}
 K_{I\bar{J}}=M^2_P{\left(\begin{array}{ccc}
 (T+\bar{T})^{-2} &  0 &  0 \\
 0 & (T_1+\bar{T}_1)^{-2}  & 0 \\
 0 &  0 &  (T_2+\bar{T}_2)^{-2}
 \end{array}\right)}
 \end{eqnarray}
for $V_{X_{i}}=0$, in which $I,J$ stand for $T,T_1,$ and $T_2$. The Kahler moduli in $K$ of Eq.~(\ref{Kahler0}) control the overall size of the compact space, 
 \begin{eqnarray}
  T=\frac{\tau}{2}+i\theta,\qquad T_i=\frac{\tau_i}{2}+i\theta_i \quad\text{with}~i=1,2\,.
 \end{eqnarray}
As can be seen from the Kahler potential above, the relevant fields participating in the four-dimensional Green-Schwarz (GS) mechanism are the $U(1)_{X_i}$ charged chiral matter superfields $\Phi_i$, the vector fields $V_{X_i}$ of the anomalous $U(1)_{X_i}$, and the Kahler moduli $T_i$. The matter superfields in $K$ consist of all the scalar fields $\Phi_i$ that are not moduli and do not have Planck sized VEVs, and the chiral matter fields $\varphi_k$ are neutral under the $U(1)_X$ symmetry. And $\delta^{\rm GS}_i$ stand for the coefficients of the mixed $U(1)_{X_i}$-$SU(3)_c$-$SU(3)_c$ color anomalies which are cancelled by the GS mechanism, $\delta^{\rm GS}_i\delta^{ab}=2\sum_{\psi_i}X_{i}\,{\rm Tr}[t^at^b]$ where $t^a$ are the generators of the representation of $SU(3)$ to which $\psi$ belongs and the sum runs over all Dirac fermions $\psi$ with $X$-charge. We take, for simplicity, the normalization factors $Z_i=Z_k=1$, and the holomorphic gauge kinetic function on the Kahler moduli 
 \begin{eqnarray}
 T_i=\frac{1}{g^2_{X_i}}+i\frac{a_{T_i}}{8\pi^2}
 \end{eqnarray}
where $g_{X_i}$ are the four-dimensional gauge couplings of $U(1)_{X_i}$. Actually, gaugino masses require a nontrivial dependence of the holomorphic gauge kinetic function on the Kahler moduli. This dependence is generic in most of the models of ${\cal N}=1$ SUGRA derived from extended supergravity and string theory~\cite{Ferrara:2011dz}. And vector multiplets $V_{X_i}$ in Eq.~(\ref{Kahler0}) are the $U(1)_{X_i}$ gauge superfields including gauge bosons $A^{\mu}_i$.
 
In the Kahler potential and superpotential in Eqs.~(\ref{Kahler0}) and (\ref{Kahler})  we have introduced two anomalous gauged $U(1)_{X}\equiv U(1)_{X_1}\times U(1)_{X_2}$ with anomalies cancelled via exchange of two Kahler-axion fields $\theta_i$ and two kinds of scalar fields $\Phi_i$ with charges $X_i$, in order to explain both the fermion mass hierarchy problem of the SM and the strong CP problem~\cite{Choi:2006qj}. The model group $SU(3)_c\times SU(2)_L\times U(1)_Y\times U(1)_{X}\times U(1)_{R}$ we are interested may be realized in a four-stack model $U(3)\times U(2)\times U(1)\times U(1)$ on D-branes where the gauged $U(1)$s are generically anomalous~\cite{string_book}. Hypercharge $U(1)_Y$ is the unique anomaly-free linear combination of the four $U(1)$s. The other combinations contribute to $U(1)_{X}$ and a gauged $U(1)_R$~\cite{Villadoro:2005yq} which contains an $R$-symmetry as a subgroup: $\{flavor\,matter\,fields\rightarrow e^{i\xi/2}flavor\,matter\,fields\}$ and $\{driving\,fields\rightarrow e^{i\xi}\,driving\,fields\}$, with $W\rightarrow e^{i\xi}W$, whereas flavon and Higgs fields remain invariant, and an axionic shift. In addition, one can introduce a non-Abelian discrete flavor symmetry, such as~\cite{Ma}, to describe flavor mixing pattern, which can be realized in field theories on orbifolds~\cite{Altarelli:2006kg}. (We will not discuss them here). $W_0$ is the constant value of the flux superpotential at its minimum. $W(T)$ is a certain non-perturbative term, which is introduced to stabilize the Kahler moduli. Although $W(T)$ in Eq.~(\ref{Kahler}) is absent at tree level, the Kahler moduli appear non-perturbatively in the superpotential through brane instantons or gaugino condensation~\cite{Derendinger:1985kk}. 
The superpotential $W_v$ dependent on the driving fields, invariant under $SU(3)_c\times SU(2)_L\times U(1)_Y\times U(1)_X\times A_4$, is given at leading order by~\cite{Ahn:2014gva}  
 \begin{eqnarray}
W_{v} &=& \Phi^{T}_{0}\big(\tilde{\mu}\,\Phi_{T}+\tilde{g}\,\Phi_{T}\Phi_{T}\big)+\Phi^{S}_{0}\big(g_{1}\,\Phi_{S}\Phi_{S}+g_{2}\,\tilde{\Theta}\Phi_{S}\big)\nonumber\\
 &+& \Theta_{0}\big(g_{3}\,\Phi_{S}\Phi_{S}+g_{4}\,\Theta\Theta+g_{5}\,\Theta\tilde{\Theta}+g_{6}\,\tilde{\Theta}\tilde{\Theta}\big)+g_{7}\,\Psi_{0}\big(\Psi\tilde{\Psi}-\mu^2_\Psi\big)\,,
 \label{potential}
 \end{eqnarray}
where $\tilde{\mu}$ is a dimensionful parameter and $\tilde{g}$, $g_{1,...,7}$ are dimensionless coupling constants. The details of the $A_{4}$ group are shown in Appendix.
The non-Abelian discrete flavor symmetry $A_{4}$ on $W_v$ is properly imposed, apart from the usual two Higgs doublets $H_{u,d}$ responsible for electroweak symmetry breaking, which are invariant under $A_{4}$ ({\it i.e.} flavor singlets), on two new types of scalar multiplets: flavon fields, responsible for the spontaneous breaking of the flavor symmetry, $\Phi_{T},\Phi_{S},\Theta,\tilde{\Theta}, \Psi, \tilde{\Psi}$ that are $SU(2)$-singlets; and driving fields $\Phi^{T}_{0},\Phi^S_{0},\Theta_{0},\Psi_{0}$ that are associated to a nontrivial scalar potential in the symmetry breaking sector. We take the flavon fields $\Phi_{T},\Phi_{S}$ to be $A_{4}$ triplets, and $\Theta,\tilde{\Theta},\Psi,\tilde{\Psi}$ to be $A_{4}$ flavor singlets, respectively, that are $SU(2)$-singlets, and driving fields $\Phi_{0}^{T},\Phi_{0}^{S}$ to be $A_{4}$ triplets and $\Theta_{0}, \Psi_{0}$ to be an $A_{4}$ singlet. In addition, there is flavored PQ symmetry $U(1)_X$ which is mainly responsible for the fermion mass hierarchy of the SM, which is composed of two anomalous {\it gauged} symmetries $U(1)_{X_1}\times U(1)_{X_2}$ generated by the charges $X_1$ and $X_2$: $\Phi_1=\{\Phi_S,\Theta,\tilde{\Theta}\}$, $\Phi_2=\{\Psi,\tilde{\Psi}\}$ are $U(1)_{X_1}$ and $U(1)_{X_2}$-charged chiral superfields, respectively. The Yukawa superpotential $W_Y$ could be appropriately arranged under the $A_4\times U(1)_{X}$ as in Ref.~\cite{Ahn:2014gva}, where the seesaw mechanism~\cite{Minkowski:1977sc} is embedded, and the fermion Yukawa couplings are visualized as functions of the gauge singlet flavon fields scaled by a cutoff proportional to string scale.

Under the $U(1)_X$ gauge transformation $V_{X_i}\rightarrow V_{X_i}+i(\Lambda_i-\bar{\Lambda}_i)$, the matter and Kahler moduli superfields transform as
 \begin{eqnarray}
  \Phi_i\rightarrow e^{iX_i\Lambda_i}\Phi_i\,,\quad T_i\rightarrow T_i+i\frac{\delta^{\rm GS}_i}{16\pi^2}\Lambda_i
 \label{transf}
 \end{eqnarray}
where $\Lambda(\bar{\Lambda}_i)$ are (anti-)chiral superfields parameterizing $U(1)_{X_i}$ transformations on the superspace. So the axionic moduli $\theta_i$ and matter axions $A_i$ have shift symmetries
 \begin{eqnarray}
  \theta_i\rightarrow \theta_i-\frac{\delta^{\rm GS}_i}{16\pi^2}\xi_i\,,\quad A_i\rightarrow A_i+X_iv_i\,\xi_i
 \label{transf_axio}
 \end{eqnarray}
where $\xi_i={\rm Re}\Lambda_i|_{\theta=\bar{\theta}=0}$ and $\Phi_i|_{\theta=\bar{\theta}=0}=\frac{1}{\sqrt{2}}e^{i\frac{A_i}{v_i}}(v_i+h_i)$ (here $v_i$ and $h_i$ being the VEVs and Higgs bosons of scalar components, respectively), with the gauge transformation 
 \begin{eqnarray}
  A^\mu_i\rightarrow A^\mu_i-\partial^\mu\xi_i\,.
 \label{transf_gauge}
 \end{eqnarray}
Then the anomaly generated by the triangle graph is cancelled by diagram in which the anomalous $U(1)_X$ mixes with the axionic moduli, which in turn couples to a multiple of the QCD instanton density ${\rm Tr}(G^{\mu\nu}\tilde{G}_{\mu\nu})$ for the gauge group in the compactification. And so the axion decay constant depends on the Kahler metric, and in particular on where the moduli are stabilized, which will be shown in Eq.~(\ref{string_axion1}).

%%%%%%%%%%%%%%%%%%%%%%%%%%%%%%%%
\section{A realistic moduli stabilization}
Since the three moduli appear in the Kahler potential Eq.~(\ref{Kahler}), by solving the $F$-term equations we are going to stabilize three size moduli with positive masses while leaving two axions massless and one axionic-partner massive through an effective superpotential $W(T)$. The two massless axion directions will be gauged by the $U(1)$ gauge interactions associated with D-branes, and the gauged flat directions of the F-term potential are removed through the Stuckelberg mechanism. The F-term scalar potential has the form 
 \begin{eqnarray}
  V_F=e^{K/M^2_P}\left\{K^{I\bar{J}}D_I W\bar{D}_{\bar{J}}\bar{W}-\frac{3}{M^2_P}|W|^2+K^{i\bar{j}}D_i W\bar{D}_{\bar{j}}\bar{W}\right\}\,,
 \end{eqnarray}
 where $K^{I\bar{J}}$ ($K^{i\bar{j}}$) is the inverse Kahler metric, and $I,J$ stand for $T,T_i$, and $i,j$ for the bosoinic components of the superfields $\Phi_i,\varphi_i$, and the Kahler covariant derivative and Kahler metric are defined as
 \begin{eqnarray}
  D_IW\equiv\partial_I W+\frac{W}{M^2_P}\partial_I K\,,\qquad K_{I\bar{J}}\equiv\partial_I\partial_{\bar{J}}K\,.
 \end{eqnarray} 
To accomplish our purpose, we take a racetrack type~\cite{Kallosh:2004yh} superpotential as an effective superpotential
 \begin{eqnarray}
  W(T)=A(\Phi_i)e^{-a(T+T_1+T_2)}+B(\Phi_i)e^{-b(T+T_1+T_2)}
 \label{WT}
 \end{eqnarray}
where $A(\Phi_i)$ and $B(\Phi_i)$ are analytic functions of $\Phi_i$ transforming under $U(1)_{X_i}$ as 
 \begin{eqnarray}
  &&A(\Phi_i)\rightarrow A(\Phi_i)\,e^{i\frac{a}{16\pi^2}(\delta^{\rm GS}_1\Lambda_1+\delta^{\rm GS}_2\Lambda_2)}\,,\nonumber\\
  &&B(\Phi_i)\rightarrow B(\Phi_i)\,e^{i\frac{b}{16\pi^2}(\delta^{\rm GS}_1\Lambda_1+\delta^{\rm GS}_2\Lambda_2)}\,,
 \label{WT_trns}
 \end{eqnarray}
 and invariant under the other gauge group. Then the scalar potential of the fields $\rho_{(i)}(=\tau_{(i)}/2)$ has local minimum at $\sigma_0,\sigma_i$ which is supersymmetric, i.e., 
 \begin{eqnarray}
  W(\sigma_0,\sigma_i)=0\,,\qquad D_TW(\sigma_0,\sigma_i)=D_{T_i}W(\sigma_0,\sigma_i)=0\,,
 \label{}
 \end{eqnarray}
 and Minkowski, i.e., $V_F(\sigma_0,\sigma_i)=0$,
 where  $\sigma_0=\sigma_i=\frac{1}{a-b}\ln\left(\frac{aA_0}{bB_0}\right)$.  And $W_0$ is fine-tuned as 
 \begin{eqnarray}
  W_0=-A_0\left(\frac{aA_0}{bB_0}\right)^{-3\frac{a}{a-b}}-B_0\left(\frac{aA_0}{bB_0}\right)^{-3\frac{b}{a-b}}\,,
 \label{W0}
 \end{eqnarray}
 where $A_0(B_0)$ are constant values of $A(\Phi_i)\left(B(\Phi_i)\right)$ at a set of VEVs $\langle\Phi_i\rangle$ that cancel all the D-terms, including the anomalous $U(1)_{X_i}$. Here the constants $W(T)$ is not analytic at the VEVs $\langle\Phi_i\rangle$ where the moduli are stabilized.
  
The F-term equations $D_TW=D_{T_i}W=0$ provide $\tau=\tau_i$, and lead to
 \begin{eqnarray}
  aA\,e^{-3\frac{a\,\tau}{2}}\,e^{-ia\,\theta^{\rm st}}+bB\,e^{-3\frac{b\,\tau}{2}}\,e^{-ib\,\theta^{\rm st}}
  +\frac{W_0+A\,e^{-3\frac{a\,\tau}{2}}\,e^{-ia\,\theta^{\rm st}}+B\,e^{-3\frac{b\,\tau}{2}}\,e^{-ib\,\theta^{\rm st}}}{\tau}=0
 \label{stabil}
 \end{eqnarray}
for $V_{X_i}=0$, where $\theta^{\rm st}\equiv\theta+\theta_1+\theta_2$. This shows that the three size moduli $(\tau,\tau_i)$ and one axionic direction $\theta^{\rm st}$ are fixed, while the two axionic directions ($\theta^{\rm st}_1\equiv\theta-\theta_1$ and $\theta^{\rm st}_2\equiv\theta-\theta_2$) are independent of the above equation. So, without loss of generality, we rebase the superfields $T$ with $\theta^{\rm st}={\rm Im}[T]$ and $T_i$ with $\theta^{\rm st}_i={\rm Im}[T_i]$ as
 \begin{eqnarray}
   T_{(i)}=\tau_{(i)}/2+i\theta_{(i)}\rightarrow T_{(i)}=\tau_{(i)}/2+i\theta^{\rm st}_{(i)}\,.
 \label{redf_K}
 \end{eqnarray}
Then  from the F-term scalar potential the masses of the fields $\rho_{(i)}$ and $\theta^{\rm st}$, $m^{2}_{\tau_{(i)}}=\frac{1}{2}K^{T\bar{T}}\partial_T\partial_{\bar{T}}V_F\Big|_{T=\bar{T}=\sigma_0}$ and $m^{2}_{\theta^{\rm st}}=\frac{1}{2}K^{T\bar{T}}\partial_{\theta^{\rm st}}\partial_{\theta^{\rm st}}V_F\Big|_{T=\bar{T}=\sigma_0}$, respectively, are obtained as follows
 \begin{eqnarray}
   m^{2}_{\tau_{(i)}}
   &=&\frac{3\ln\Big(\frac{aA_0}{bB_0}\Big)}{M^4_P(a-b)}\Big\{A_0a^2\Big(\frac{aA_0}{bB_0}\Big)^{-3\frac{a}{a-b}}+B_0b^2\Big(\frac{aA_0}{bB_0}\Big)^{-3\frac{b}{a-b}}\Big\}^2\,,\\
 m^{2}_{\theta^{\rm st}}
 &=&\frac{3W_0}{M^4_P}\Big\{-A_0a^3\Big(\frac{aA_0}{bB_0}\Big)^{-3\frac{a}{a-b}}-B_0b^3\Big(\frac{aA_0}{bB_0}\Big)^{-3\frac{b}{a-b}}\Big\}\nonumber\\
 &+&\frac{6\ln\left(\frac{aA_0}{bB_0}\right)}{M^4_P(a-b)}\Big\{-A_0B_0(a-b)^2\Big(\frac{aA_0}{bB_0}\Big)^{-3\frac{a+b}{a-b}}\Big(\frac{a^2-b^2}{2\ln\big(\frac{aA_0}{bB_0}\big)}+ab\Big)\Big\}\,.
 \label{massT}
 \end{eqnarray}
Here the mass squared of the size moduli fields $\rho_{(i)}$ at the minimum is simply given by $m^{2}_{\tau_{(i)}}=3\sigma_0\left|W_{TT}(\sigma_0)\right|^2/M^4_P$ where $W_{TT}=\partial^2W/(\partial T)^2$. Note that the gravitino mass in this supersymmetric Minkowski minimum vanishes. 
With the conditions $a<0$, $b>0$ ($|a|<|b|$) and $A_0>0$, $B_0<0$ we obtain positive values of masses.
Here $a,b$ are constants, while $A_0,B_0$ are constants in $M^3_P$ units. For a simple choice of parameters, $A_0=-B_0=0.01$,  $a=-2\pi/100$ and $b=2\pi/90$, one has $m_\tau\simeq1.7\times10^{14}$ GeV and $m_{\theta_{\rm st}}\simeq2.0\times10^{14}$ GeV.

%%%%%%%%%%%%%%%%%%%%%%%%%%%%%%%%
\section{Supersymmetry breaking}
As discussed in the Kallosh-Linde model~\cite{Kallosh:2004yh}, supersymmetry is unbroken so far in the vacuum states corresponding to the minimum of the potential with $V=0$.
 As will be shown later, the existence of Fayet-Iliopoulos (FI) terms $\xi^{\rm FI}_i$ for the corresponding $U(1)_{X_i}$ implies the existence of uplifting potential which makes a nearly vanishing cosmological constant and induces SUSY breaking. A small perturbation $\Delta W$ to the superpotential~\cite{Kallosh:2004yh, Kallosh:2003} is introduced in order to determine SUSY breaking scale. Then the minimum of the potential is shifted from zero to a slightly negative value at $\sigma_0+\delta\rho$, $\sigma_i+\delta\rho_i$ by the small constant $\Delta W$. The resulting F-term potential has a supersymmetric Anti de Sitter(AdS) minimum and consequently the depth of this minimum is given in terms of $W(\sigma_0+\delta\rho, \sigma_i+\delta\rho_i)\simeq\Delta W+{\cal O}(\Delta W)^2$ by
 \begin{eqnarray}
  V_{\rm AdS}\simeq-\frac{3}{M^2_P}\frac{(\Delta W)^2}{8\sigma_0\sigma_1\sigma_2}=-\frac{3}{8M^2_P}\Big(\frac{a-b}{\ln\frac{aA_0}{bB_0}}\Big)^2(\Delta W)^2\,,
 \label{AdS}
 \end{eqnarray}
 where $\Delta W=\langle W\rangle_{\rm AdS}$ is the value of the superpotential at the AdS minimum.
At the shifted minimum SUSY is preserved, {\it i.e.} $D_TW(\sigma_0+\delta\rho)=0$ and $D_{T_i}W(\sigma_i+\delta\rho_i)=0$, leading to $W_T(\sigma_0+\delta\rho)=W_{T_i}(\sigma_0+\delta\rho_i)\simeq3\Delta W/2\sigma_0$.
At this new minimum the displacements $\delta\rho=\delta\rho_{i}$ are obtained as
 \begin{eqnarray}
  \delta\rho_{(i)}\simeq\frac{3\Delta W}{2\sigma_0 W_{TT}(\sigma_0)}
  = \frac{3(a-b)\Delta W}{2\ln\left(\frac{aA_0}{bB_0}\right)\Big\{A_0a^2\Big(\frac{aA_0}{bB_0}\Big)^{\frac{-3a}{a-b}}+B_0b^2\Big(\frac{aA_0}{bB_0}\Big)^{\frac{-3b}{a-b}}\Big\}}\,.
 \label{}
 \end{eqnarray}
After adding the uplifting potentials SUSY is broken and then the gravitino in the uplifted minimum acquires a mass
 \begin{eqnarray}
  m_{3/2}\simeq\frac{|\Delta W|}{M^2_P}\Big(\frac{a-b}{2\ln\frac{aA_0}{bB_0}}\Big)^{\frac{3}{2}}\,.
 \label{gravitino}
 \end{eqnarray}
The uplifting of the AdS minimum to the dS minimum can be achieved by considering non-trivial fluxes for the gauge fields living on the D7 branes~\cite{Burgess:2003ic} which can be identified as field-dependent FI D-terms in the ${\cal N}=1$, $4D$ effective action~\cite{Brunner:1999jq}.
The uplifting terms can be parameterized as $\Delta V_i=\frac{1}{2}(\xi^{\rm FI}_i)^2g^2_{X_i}\simeq|V_{\rm AdS}|(\sigma_0/\rho_i)^3$~\cite{Burgess:2003ic} such that the value of the potential at the new minimum become equal to the observed value of the cosmological constant.  So, as will be shown later, the anomalous FI terms can not be cancelled, and act as uplifting potential. And expanding the Kahler potential $K$ in components, the term linear in $V_{X_i}$ produces the FI factors $\xi^{\rm FI}_i=\frac{\partial K}{\partial V_{X_i}}\big|_{V_{X_i}=0}\,\Delta\rho_i$ as
 \begin{eqnarray}
  \xi^{\rm FI}_i=M^2_P\frac{\delta^{\rm GS}_i}{16\pi^2\tau}\Delta\rho_i\,.
 \label{FI}
 \end{eqnarray}
Here the displacements $\Delta\rho_i\equiv\rho_i-\sigma_0$ in the moduli fields are induced by the uplifting terms, 
 \begin{eqnarray}
  \Delta\rho_i\simeq\frac{6M^2_P|V_{\rm AdS}|}{W^2_{TT}(\sigma_0)}\,,
 \end{eqnarray}
which are achieved by $\partial_{\rho_i}(V_F+\Delta V_i)|_{\sigma_i+\delta\rho_i}=0$. Since the uplifting terms by $\Delta\rho_i$ making the dS minimum induce SUSY breaking, all particles whose mass is protected from supersymmetry become massive. 
With our choice of parameters, the gravitino mass being of order $10$ TeV implies $|\Delta W|\simeq10^{-14}M^3_P$, and which in turn means that the FI terms proportional to $|V_{\rm AdS}|/m^2_\tau$ are expected to be strongly suppressed.

Setting to zero from the beginning the SM matter fields $\{q^c,\ell,H_u,...\}$, with the almost vanishing cosmological constant for the remaining fields the gravitino mass $m_{3/2}$ is directly related to the scale of supersymmetry breaking, $|F|^2-3m^2_{3/2}M^2_P+D^2_{X_i}/2\approx0$, implying that the F- and D-term potentials should vanish in the limit $m_{3/2}$ going to zero and some of them should scale as $m_{3/2}$ at the minimum. 
In the global SUSY limit, {\it i.e.} $M_{P}\rightarrow\infty$,  the relevant F-term potential is written as
\begin{eqnarray}
 V^{\rm global}_{F}&=&\left|\frac{2g_{1}}{\sqrt{3}}\left(\Phi_{S1}\Phi_{S1}-\Phi_{S2}\Phi_{S3}\right)+g_{2}\Phi_{S1}\tilde{\Theta}\right|^{2}\nonumber\\
  &+&\left|\frac{2g_{1}}{\sqrt{3}}\left(\Phi_{S2}\Phi_{S2}-\Phi_{S1}\Phi_{S3}\right)+g_{2}\Phi_{S3}\tilde{\Theta}\right|^{2}\nonumber\\
  &+&\left|\frac{2g_{1}}{\sqrt{3}}\left(\Phi_{S3}\Phi_{S3}-\Phi_{S1}\Phi_{S2}\right)+g_{2}\Phi_{S2}\tilde{\Theta}\right|^{2}\nonumber\\
  &+&\left|g_{3}\left(\Phi_{S1}\Phi_{S1}+2\Phi_{S2}\Phi_{S3}\right)+g_{4}\Theta^{2}+g_{5}\Theta\tilde{\Theta}+g_{6}\tilde{\Theta}^{2}\right|^{2}\nonumber\\
  &+&\left|g_{7}\left(\Psi\tilde{\Psi}-\mu^2_\Psi\right)\right|^2+|g_7|^2|\Psi_0|^2\left(|\Psi|^2+|\tilde{\Psi}|^2\right)+\sum_{i={\rm the~others}}\left|\frac{\partial W_{v}}{\partial z_{i}}\right|^{2}\,,
 \label{V_F}
\end{eqnarray}
and the D-term potential, obtained by the introduction of two FI D-terms ${\cal L}^{\rm FI}_i=-g_{X_i}\xi^{\rm FI}_i D_{X_i}$, is given by
 \begin{eqnarray}
  V^{\rm global}_D&=&\frac{|X_1|^2g^2_{X_1}}{2}\Big(\frac{\xi^{\rm FI}_1}{|X_1|}-|\Phi_S|^2-|\Theta|^2-|\tilde{\Theta}|^2\Big)^2
  +\frac{|X_2|^2g^2_{X_2}}{2}\Big(\frac{\xi^{\rm FI}_2}{|X_2|}-|\Psi|^2+|\tilde{\Psi}|^2\Big)^2
 \label{grobal_V}
 \end{eqnarray}
with $D_{X_i}=g_{X_i}(\xi^{\rm FI}_i-\sum_i X_i|\Phi_i|^2)$, where $\xi^{\rm FI}_i=2E_i/\tau_i$ are constant parameters with dimensions of mass squared and here $E_i$ are measure of the strength of the fluxes for the gauge fields living on the D7 branes~\cite{Burgess:2003ic} . 
Since SUSY is preserved after the spontaneous symmetry breaking of $U(1)_X\times A_4$, the scalar potential in the limit $M_P\rightarrow\infty$ vanishes at its ground states, {\it i.e.}, vanishing F-terms must have also vanishing D-terms. 
Consequently, the VEVs of the flavon fields are from the minimization conditions of the F-term scalar potential: the phenomenologically non-trivial solutions~\cite{Ahn:2014gva}  
 \begin{eqnarray}
  \langle\Phi_S\rangle=\frac{1}{\sqrt{2}}(v_S,v_S,v_S)\,, \qquad\langle\Theta\rangle=\frac{v_\Theta}{\sqrt{2}}\,, \qquad\langle\Psi\rangle=\langle\tilde{\Psi}\rangle=\frac{v_\Psi}{\sqrt{2}}\,,
 \label{vev_ph}
 \end{eqnarray}
as well as a set of trivial solutions
 \begin{eqnarray}
  \langle\Phi_S\rangle=(0,0,0)\,, \qquad\langle\Theta\rangle=0\,, \qquad\langle\Psi\rangle=\langle\tilde{\Psi}\rangle=\frac{v_\Psi}{\sqrt{2}}\,.
 \label{vev_tr}
 \end{eqnarray}
Then the two supersymmetric solutions are taken by the D-flatness conditions, respectively, for (i) phenomenologically viable case 
 \begin{eqnarray}
  \xi^{\rm FI}_1=|X_1|(\langle|\Phi_S|^2\rangle+\langle|\Theta|^2\rangle)\,, \qquad\xi^{\rm FI}_2=0\,, \qquad\langle\Psi\rangle=\langle\tilde{\Psi}\rangle\,,
 \end{eqnarray}
 and (ii) phenomenologically trivial case 
 \begin{eqnarray}
  \xi^{\rm FI}_1=\langle\Phi_S\rangle=\langle\Theta\rangle=0\,, \qquad\xi^{\rm FI}_2=0\,, \qquad\langle\Psi\rangle=\langle\tilde{\Psi}\rangle\,,
 \end{eqnarray}
both of which indicate that the VEVs of the flavon fields strictly depend on the moduli stabilization, particularly on the VEVs of the fluxes $E_i$ in the FI terms~\cite{Burgess:2003ic}. 
So it seems hard for the first case (i) to stabilize $|\Phi_i|$ at large VEVs$\sim{\cal O}(10^{12})$ GeV. And there is a tension between $\langle\Phi_i\rangle=0$ and $\langle\xi^{\rm FI}_i\rangle\neq0$ which are possible as long as $E_i$ are below the string scale. Therefore it is imperative that, in order for the D-terms to act as uplifting potential, the F-terms have to necessarily break SUSY. 
In order for the solution in Eq.~(\ref{vev_tr})  to be phenomenologically non-trivial, we destabilize $\Phi_1=\{\Phi_S,\Theta\}$ and $\Phi_2=\{\Psi,\tilde{\Psi}\}$ by their tachyonic SUSY masses to develop $v_S$, $v_\Theta$, $v_{\Psi}(v_{\tilde{\Psi}})$ comparable with seesaw and QCD axion window scales~\cite{Ahn:2014gva}, while keeping $\langle\tilde{\Theta}\rangle=0$ for the scalar field $\tilde{\Theta}$ with $m^2_{\tilde{\Theta}}>0$. The phenomenologically viable VEVs of the flavon fields can be determined by considering both the SUSY breaking effect which lift up the flat directions and supersymmetric next-to-leading order terms (see the origin of this argument~\cite{Murayama:1992dj}) invariant under $A_4\times U(1)_X$. 
The supersymmetric next-to-leading order terms are given by
 \begin{eqnarray}
  \Delta W_v&\simeq&\frac{\alpha}{m_{P}}\Psi\tilde{\Psi}(\Phi_T\Phi^T_0)_{\bf 1}+\frac{\beta}{m_P}(\Phi^S_0\Phi_T)_{\bf 1}\Theta\Theta\nonumber\\
  &+&\frac{1}{m_P}\left\{\gamma_1(\Phi_S\Phi_S)_{\bf 1}(\Phi_T\Phi^S_0)_{\bf 1}+\gamma_2(\Phi_S\Phi_S)_{\bf 1'}(\Phi_T\Phi^S_0)_{\bf 1''}+\gamma_3(\Phi_S\Phi_S)_{\bf 1''}(\Phi_T\Phi^S_0)_{\bf 1'}\right\}\,,
 \end{eqnarray}
where $\alpha$, $\beta$, and $\gamma_{1,2,3}$ are real-valued constants being of order unities. Note that here, since we are considering the phenomenologically non-trivial solutions as in Eq.~(\ref{vev_ph}), operators including $\tilde{\Theta}$, $(\Phi_S\Phi_S)_{\bf 3s}$, $(\Phi_S\Phi_T)_{\bf 3s}$, and $(\Phi_S\Phi_T)_{\bf 3a}$ are neglected in $\Delta W_v$.
Since soft SUSY-breaking terms are already present at the scale relevant to flavor dynamics, the scalar potentials for $\Psi(\tilde{\Psi})$ and $\Phi_S(\Theta)$ at leading order read
\begin{eqnarray}
  V(\Phi_S,\Theta)&\simeq&\beta_1m^2_{3/2}|\Phi_S|^2+\beta_2m^2_{3/2}|\Theta|^2+\frac{v_T^2|\beta\Theta^2+\gamma\Phi^2_S|^2}{2m^2_{P}},\nonumber\\
  V(\Psi,\tilde{\Psi})&\simeq&\alpha_1m^2_{3/2}|\Psi|^2+\alpha_2m^2_{3/2}|\tilde{\Psi}|^2+|\alpha|^2\frac{v_T^2|\Psi|^2|\tilde{\Psi}|^2}{2m^2_{P}},
 \label{V_susy}
 \end{eqnarray}
leading to the PQ breaking scales
 \begin{eqnarray}
  \mu^2_{\Psi}&=&\frac{v_{\Psi}v_{\tilde{\Psi}}}{2}=\frac{2\sqrt{\alpha_1\alpha_2}}{|\alpha|^2}\,\left(\frac{m_{3/2}}{v_T}\,m_{P}\right)^{2}\,,\\
 v^2_S&=& \frac{2\,\beta_1\,\kappa^2}{\gamma\,(\beta+\gamma)}\,\left(\frac{m_{3/2}}{v_T}\,m_{P}\right)^2=\kappa^2\,v^2_\Theta\,,
 \label{PQ_scale}
\end{eqnarray}
where $\gamma=3(\gamma_1+\gamma_2+\gamma_3)$, $\beta_1\beta=\gamma\beta_2$, and $\kappa=(-3g_3/g_4)^{-\frac{1}{2}}$.  It indicates that the gravitino mass (or SUSY breaking mass) strongly depends on the scale of PQ fields as well as $\Phi_T$; for example, for $\mu_\Psi\sim10^{13}$ GeV and $v_T\sim10^{11}$ GeV satisfying the SM fermion mass hierarchies~\cite{Ahn:2014gva} one can obtain $m_{3/2}\sim{\cal O}(10)$ TeV, and/or subsequently $v_S\sim v_\Theta\sim10^{11}$ GeV.
With the soft SUSY-breaking potential, the radial components of the fields $\Psi$ and $\tilde{\Psi}$ are stabilized at
\begin{eqnarray}
  v_{\Psi}\simeq\mu_{\Psi}\sqrt{2}\left(\frac{\alpha_2}{\alpha_1}\right)^{1/4}\,,\qquad v_{\tilde{\Psi}}\simeq\mu_{\Psi}\sqrt{2}\left(\frac{\alpha_1}{\alpha_2}\right)^{1/4}\,,
 \label{PQ_scale1}
\end{eqnarray}
respectively.

%%%%%%%%%%%%%%%%%%%%%%%%%%%%%%%%
\section{String inspired QCD axions}
Finally, we consider the four-dimensional effective Lagrangian of the axions, $\theta^{\rm st}_i$ and $A_i$, and the $U(1)_X$ gauge fields, $A^{\mu}_i$, which contains the following
%\begin{widetext}
  \begin{eqnarray}
  &&K_{T_i\bar{T}_i}\Big(\partial^{\mu}\theta^{\rm st}_i-\frac{\delta^{\rm GS}_{i}}{16\pi^2}A^{\mu}_i\Big)^2-\frac{1}{4g^2_{X_i}}F^{\mu\nu}_iF_{i\mu\nu}-g_{X_i}\xi^{\rm FI}_iD_{X_i}\nonumber\\
  &&+|D_\mu\Phi_i|^2+\theta^{\rm st}_i{\rm Tr}(G^{\mu\nu}\tilde{G}_{\mu\nu})+\frac{A_i}{X_iv_i}\frac{\delta^{\rm GS}_i}{16\pi^2}{\rm Tr}(G^{\mu\nu}\tilde{G}_{\mu\nu})\,,
 \label{string_axion}
 \end{eqnarray}
 %\end{widetext}
where $F^{\mu\nu}_i$ are the $U(1)_{X_i}$ gauge field strengths $F^{\mu\nu}_i=\partial^\mu A^\nu_i-\partial^\nu A^\mu_i$, and the QCD gauge couplings are absorbed into the gluon field strengths.
In $|D_\mu\Phi_i|^2$ the scalar fields $\Phi_i$ couple to the $U(1)_{X_i}$ gauge bosons, where the gauge couplings $g_{X_i}$ are absorbed into the gauge bosons $A^{\mu}_i$ in the $U(1)_X$ gauge covariant derivative $D^\mu\equiv\partial^\mu+iX_iA^\mu_i$. As mentioned before, the introduction of FI terms ${\cal L}_{\rm FI}=-\xi^{\rm FI}_i\int d^2\theta V_{X_i}=-\xi^{\rm FI}_ig_{X_i}D_{X_i}$ leads to the D-term potentials in Eq.~(\ref{grobal_V}) where the FI factors $\xi^{\rm FI}_i$ depend on the closed string moduli $\rho_i=\tau_i/2$. The first, third and fourth terms of Eq.~(\ref{string_axion}) stem from expanding the Kahler potential of Eq.~(\ref{Kahler0}).
Under the anomalous $U(1)_X$ gauge transformation in Eqs.(\ref{transf}) and (\ref{transf_axio}), the first and fifth terms together, and similarly the fourth and sixth terms in Eq.~(\ref{string_axion}), are gauge invariant, that is, the interaction Lagrangians 
  \begin{eqnarray}
  {\cal L}^{\rm int}_{X_i}&=&A^{\mu}_iJ^{X_i}_\mu-\frac{A_i}{X_iv_i}\frac{\delta^{\rm GS}_i}{16\pi^2}\,{\rm Tr}(G^{\mu\nu}\tilde{G}_{\mu\nu})\,,\nonumber\\
  {\cal L}^{\rm int}_{\theta^{\rm st}_i}&=&A^{\mu}_iJ^{X_i}_\mu+\theta^{\rm st}_i\,{\rm Tr}(G^{\mu\nu}\tilde{G}_{\mu\nu})\,,
 \label{}
 \end{eqnarray}
are invariant. There are anomalous currents $J^{X_i}_\mu$ coupling to the gauge bosons $A^\mu_{i}$, that is, $\partial_\mu J^{\mu}_{X_i}=\frac{\delta^{\rm GS}_i}{16\pi^2}\,{\rm Tr}(G^{\mu\nu}\tilde{G}_{\mu\nu})$:
 \begin{eqnarray}
  J^{X_i}_\mu=K_{T_i\bar{T}_i}\frac{\delta^{\rm GS}_i}{8\pi^2}\partial_\mu\theta^{\rm st}_i-iX_i\Phi^\ast_i\overleftrightarrow{\partial_\mu}\Phi_i
  +\frac{1}{2}\sum_{\psi_i}X_{i}\bar{\psi}_i\gamma_\mu\gamma_5\psi_i\,.
 \label{X_current}
 \end{eqnarray}
 Expanding Lagrangian (\ref{string_axion}) and using $\theta^{\rm st}_{i}=a_{T_i}/8\pi^2$ it reads
  \begin{eqnarray}
  &&\frac{1}{2}\left(\partial^{\mu}\tilde{a}_{T_i}\right)^2+\frac{\tilde{a}_{T_i}}{f^{\rm st}_i}\frac{1}{8\pi^2}{\rm Tr}(G^{\mu\nu}\tilde{G}_{\mu\nu})+\frac{1}{2}\left(\partial^{\mu}A_i\right)^2+\frac{A_i}{X_iv_i}\frac{\delta^{\rm GS}_i}{16\pi^2}{\rm Tr}(G^{\mu\nu}\tilde{G}_{\mu\nu})\nonumber\\
  &&-J^{X_i}_\mu A^\mu_i+\frac{1}{2g^2_{X_i}}m^2_{X_i}A^\mu_iA_{i\mu}-\frac{1}{4g^2_{X_i}}F^{\mu\nu}_iF_{i\mu\nu}-\frac{g^2_{X_i}}{2}\Big(\xi^{\rm FI}_i-\sum_i X_i|\Phi_i|^2\Big)^2
 \label{string_axion1}
 \end{eqnarray}
 where $\tilde{a}_{T_i}=f^{\rm st}_ia_{T_i}$ is the canonically normalized Kahler axions with $f^{\rm st}_i=\sqrt{2K_{T_i\bar{T}_i}/(8\pi^2)^2}$. Clearly it indicates that the values of $f^{\rm st}_i$ depend on the Kahler metric and where on the moduli are stabilized. And the gauge boson masses obtained by the Higgs mechanism are given by
  \begin{eqnarray}
  m_{X_i}=g_{X_i}\sqrt{2K_{T_i\bar{T}_i}\Big(\frac{\delta^{\rm GS}_i}{16\pi^2}\Big)^2+2f^2_{\Phi_i}} \,,
 \label{mX_mass}
 \end{eqnarray}
Then the open string axions $A_i$ are linearly mixed with the closed string axions $\tilde{a}_{T_i}$ with decay constants $f^{\rm st}_i$ and $f_{\Phi_i}=X_iv_i$
 \begin{eqnarray}
  \tilde{A}_i=\frac{A_i\frac{\delta^{\rm GS}_i}{2}f^{\rm st}_i-\tilde{a}_{T_i}f_{\Phi_i}}{\sqrt{f^2_{\Phi_i}+(\frac{\delta^{\rm GS}_i}{2}f^{\rm st}_i)^2}}\,,\qquad
  G_i=\frac{\tilde{a}_{T_i}\frac{\delta^{\rm GS}_i}{2}f^{\rm st}_i+A_if_{\Phi_i}}{\sqrt{f^2_{\Phi_i}+(\frac{\delta^{\rm GS}_i}{2}f^{\rm st}_i)^2}}\,.
 \label{axion1}
 \end{eqnarray}
Since the $U(1)_X$ is gauged, two linear combinations $G_i$ of the fields $A_i$ and $\tilde{a}_{T_i}$ are eaten by the $U(1)_X$ gauge bosons and obtain string scale masses, while the other combinations $\tilde{A}_i$ survive to low energies and contribute to the QCD axion. With the given parameters we obtain $m_{X_1}\sim10^{16}$ GeV and $m_{X_2}\sim10^{17}$ GeV for $\tau_i/2\sim1$, $\delta^{\rm GS}_1=3$, and $\delta^{\rm GS}_2=17$.
For $f^{\rm st}_i\gg v_i$, the axions $\tilde{A}_i$ as would-be QCD axion are approximated to $A_i$. Below the scale $m_{X_i}$ the gauge bosons decouple, leaving behind low-energy symmetries which are anomalous global $U(1)_{X_i}$. One linear combination of the global $U(1)_{X_i}$ is broken explicitly by instantons. The rigid example of such would-be QCD axions, and some of its consequences were studied in Ref.~\cite{Ahn:2014gva}. See also Refs.~\cite{Honecker:2013mya}. 

%%%%%%%%%%%%%%%%%%%%%%%%%%%%%%%%
\section{Conclusion}
We constructed a string-inspired model as a useful bridge between flavor physics and string theory by introducing two anomalous gauged $U(1)$ symmetries responsible for both the fermion mass hierarchy problem of the SM and the strong CP problem. In the context of supersymmetric moduli stabilization we strongly stabilized the size moduli with positive masses while leaving two axions massless and one axion massive. We showed that, while the massive gauge bosons eat the two axionic degrees of freedom, two axionic directions survive to low energies  as the flavored PQ axions.

\newpage
\appendix
%%%%%%%%%%%%%%%%%%%%%%%%%%%%%%%%%%%%%%%%%%%%%%%%%%%%%%%%%%%%%%%%%%%%
\section{Non-Abelian discrete symmetry $A_4$}
The group $A_{4}$ is the symmetry group of the tetrahedron and the finite groups of the even permutation of four objects having four irreducible representations: its irreducible representations are one triplet ${\bf 3}$ and three singlets ${\bf 1}, {\bf 1}', {\bf 1}''$
with the multiplication rules ${\bf 3}\otimes{\bf 3}={\bf 3}_{s}\oplus{\bf 3}_{a}\oplus{\bf 1}\oplus{\bf 1}'\oplus{\bf 1}''$,
${\bf 1}'\otimes{\bf 1}''={\bf 1}$, ${\bf 1}'\otimes{\bf 1}'={\bf 1}''$
and ${\bf 1}''\otimes{\bf 1}''={\bf 1}'$.
Let $(a_{1}, a_{2}, a_{3})$ and $(b_{1}, b_{2}, b_{3})$ denote the basis vectors for two ${\bf 3}$'s. Then, we have
 \begin{eqnarray}
  (a\otimes b)_{{\bf 3}_{\rm s}} &=& \frac{1}{\sqrt{3}}(2a_{1}b_{1}-a_{2}b_{3}-a_{3}b_{2}, 2a_{3}b_{3}-a_{2}b_{1}-a_{1}b_{2}, 2a_{2}b_{2}-a_{3}b_{1}-a_{1}b_{3})~,\nonumber\\
  (a\otimes b_c)_{{\bf 3}_{\rm a}} &=& i(a_{3}b_{2}-a_{2}b_{3}, a_{2}b_{1}-a_{1}b_{2}, a_{1}b_{3}-a_{3}b_{1})~,\nonumber\\
  (a\otimes b)_{{\bf 1}} &=& a_{1}b_{1}+a_{2}b_{3}+a_{3}b_{2}~,\nonumber\\
  (a\otimes b)_{{\bf 1}'} &=& a_{1}b_{2}+a_{2}b_{1}+a_{3}b_{3}~,\nonumber\\
  (a\otimes b)_{{\bf 1}''} &=& a_{1}b_{3}+a_{2}b_{2}+a_{3}b_{1}~.
  \label{A4reps}
 \end{eqnarray}
%%%%%%%%%%%%%%%%%%%%%%%%%%%%%%
\acknowledgments{This work was supported by IBS under the project code, IBS-R018-D1. We thank Stephen Angus for helpful discussions.
}

%%%%%%%%%%%%%%%%%%%%%%%%%%%%%%%%%%%%%%%%%%%%%%%%%%%%%%%%%%%%%%%%%%%%%%%%%%%%%%%%%%%%%

\end{document}